\documentclass[12pt]{article}

\newread\testifexists
\def\GetIfExists #1 {\immediate\openin\testifexists=#1
    \ifeof\testifexists\immediate\closein\testifexists\else
    \immediate\closein\testifexists\input #1\fi}

\usepackage{gthstyle}
\usepackage{amsfonts}
\usepackage{amssymb}
\immediate\openin\testifexists=e
    \ifeof\testifexists\immediate\closein\testifexists\else
    \immediate\closein\testifexists\input e\fipsf
\def\Bbb#1{\setbox0=\hbox{$\tt #1$}  \copy0\kern-\wd0\kern .1em\copy0}
\def\bbf#1{\setbox0=\hbox{$#1$} \kern-.025em\copy0\kern-\wd0
        \kern.05em\copy0\kern-\wd0 \kern-.025em\raise.0433em\box0}
\def\ffract#1#2{\raise .3 em\hbox{$\scriptstyle#1$}\kern-.25em/
                \kern-.2em\lower .2 em \hbox{$\scriptstyle#2$}}

\def\part#1#2{{\partial#1\over\partial#2}}

\newcommand{\be}{\begin{eqnarray}}
\newcommand{\ee}{\end{eqnarray}}

\newcommand{\bi}[1]{\begin{itemize}\item[#1]}

\newcommand{\ei}{\end{itemize}}

\newcommand{\fn}{\footnote}

\newcommand{\newsec}[1]{\section{#1}\setcounter{equation}{0}}
\def\printversion{\setlength{\textheight}{9in}\setlength{\oddsidemargin}{0in}
    \setlength{\textwidth}{6.3in}\setlength{\topmargin}{-0.1in}}

\newcommand {\eel}[1]{\label{#1}\end{eqnarray}} % equationnumbers

\printversion 
\begin{document} \begin{titlepage}

\title{\normalsize \hfill ITP-UU-07/38  \\ \hfill SPIN-07/26
\\ \hfill {\tt hep-th/yymmnnn}\\ \vskip 20mm \Large\bf
THE GRAND VIEW OF PHYSICS \thanks{Presented at \emph{Salam +50}, Imperial College, London, July 7, 2007}}

\author{Gerard 't~Hooft}
\date{\normalsize Institute for Theoretical Physics \\
Utrecht University \\ and
\medskip \\ Spinoza Institute \\ Postbox 80.195 \\ 3508 TD
Utrecht, the Netherlands \smallskip \\ e-mail: \tt g.thooft@phys.uu.nl \\ internet: \tt http://www.phys.uu.nl/\~{}thooft/}

\maketitle

\begin{quotation} \noindent {\large\bf Abstract } \medskip \\ Abdus Salam was known for his `grand views', grand
views of science as well as grand views of society. In this talk the grand view of theoretical physics is put
in perspective.
\end{quotation}

\vfill \flushleft{\today}

\end{titlepage}

\eject
\newsec{Confronting challenges}
To obtain the Grand Picture of the physical world we inhabit, to identify the real problems and distinguish
them from technical details, to spot the very deeply hidden areas where there is room for genuine improvement
and revolutionary progress, courage is required. Every now and then, one has to take a step backwards, one
has to ask silly questions, one must question established wisdom, one must play with ideas like being a
child. And one must not be afraid of making dumb mistakes.

By his adversaries, Abdus Salam was accused of all these things. He could be a child in his wonder about
beauty and esthetics, and he could make mistakes. Glancing back at his numerous and wildly varied
publications, I can see why some people found it difficult to understand why he was given the most
prestigious award of our trade, the Nobel Prize, since even the one publication that is cited most, his work
with John C. Ward, places the leptons in a multiplet that no-one today would find acceptable, and even in his
days, it could easily be argued why that proposition would have to fail. Salam himself was also surprised by
the quotation that went with his Nobel Prize. ``I think", he once confided to me, ``that the Nobel Committee
also rewarded me for my ideas about the two-component neutrino. That was right on the spot, and I was the
first".

But that was not what he was rewarded for. I am sure that the Nobel Committee would have mentioned it.
Instead, the committee stated: \begin{quotation}\noindent``for their contributions to the theory of the
unified weak and electromagnetic interaction between elementary particles, including, inter alia, the
prediction of the weak neutral current".\end{quotation} Indeed, the prediction of the existence of a neutral
component in the weak current happened to be correctly predicted even if you took the wrong multiplet. The
real reason why he earned the Nobel was that he had the grand view. As Sidney Coleman had noticed, ``Salam
even gave the essence of a correct argument for his belief in the renormalizability of the theory". And, one
cannot emphasize enough, \emph{that} is what counts. After all, we now know that what is called the
``Standard Model" today, is just one apparently haphazard choice, made by Nature, among many possibilities of
principle. Salam's multiplet \emph{could} have been right; his grand vision certainly was right.

This grand view evolved extensively. The principle that truly dominated the later half of the 20th century
was the principle of symmetry. ``If you can identify Nature's complete symmetry group, you will know
everything", is what became a pivotal dogma. But before this insight was truly appreciated, one first had to
overcome a major obstacle: divergences. Are the divergences in the integration expressions for quantized
field theories a fundamental shortcoming of the general idea, or can they be overcome, so that the
divergences can be viewed as nothing more than a temporary technical obstacle? Salam had been strongly
attracted to this question. But indeed he posed the question, rather than airing dogmatic views on the
subject, like so many of his contemporaries.

Did he see it right? Salam did notice that overlapping divergences can be disentangled, that divergences in
many theories can indeed be seen as a technical problem that does not disqualify quantized field theories as
a whole, and that exact gauge symmetry is essential for the theories to work. But then he played and made
mistakes. Any attempt to quantize gravity is also beset by divergences. Well, having put aside the divergence
problem as a technical one, perhaps it is a technical problem in gravity as well? When I met him frequently,
he was fiercely attempting to rearrange the Feynman diagrams of quantum gravity. Since the interaction is
non-polynomial (in most gauges, that is), one can observe that any pair of vertices in gravity could be
connected by an indefinite number of propagators, the \emph{superpropagator} (see Fig.~\ref{superprop.fig}

\begin{figure}[h] \setcounter{figure}{0}
\begin{quotation}
 \epsfxsize=65 mm\epsfbox{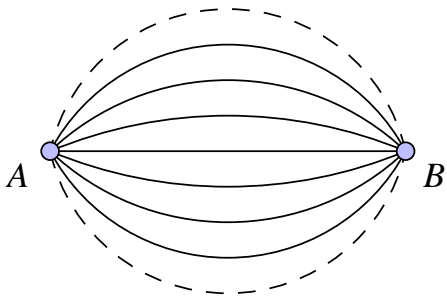}
  \caption{The superpropagator. Two vertices, \(A\) and \(B\), are connected by an indefinite number of
  elementary propagators.}
  \label{superprop.fig}\end{quotation}
\end{figure}

As many other attempts to tame quantum gravity, this one was bound to fail. Here I never could agree with
Salam, for very fundamental reasons. My problem with this is the same difficulty I have with claims from both
supergravity and string theory today. Just imagine that indeed the integrals that one needs to perform would
converge, or at least become sufficiently regular, whenever the momentum variables $k_\mu$ would tend to
infinity. This would mean that some kind of decoupling would take place at very high momentum, or
equivalently, at very tiny distance scales. This decoupling would indicate that, at tiny distance scales, our
system would linearize, decouple, simplify. We would be able to describe a smooth and comprehensible world at
distance scales smaller than, say, the Planck length. Why not try to imagine such a world directly? The point
is that this is impossible. Newton's constant would tend to infinity there, or, distances in space and time,
as what we are familiar with, cease to make sense there. This is characteristic for a \emph{topological}
theory. Thus, gravity must become purely topological at small distances. As long as we do not have such a
topological theory, chances that we stumble upon one by blindly manipulating superpropagators, supergravity
diagrams or string world sheets, are remote. Our searches should be well directed ones.

\newsec{Grand Unification}
Should the color group SU(3) be extended to SU(4) so as to ``unify" the leptons with the quarks? This is what
Salam thought, and he further pursued the thought with Jogesh Pati, ideas that were also expanded by Georgi,
Glashow, Quinn and De Rujula. It so turned out that the fermions most naturally fit in a \(10\oplus \overline
5\) representation of SU(5), a group that combined color SU(3) with electro-weak SU(2)\(\times\)U(1). It
later turned out that this SU(5) did not quite agree with observations: it would predict a too rapid proton
decay. Extending SU(5) to SO(10) is very straightforward. The \(10\oplus \overline 5\) combine with a SU(5)
singlet into a 16. The singlet was the right-handed neutrino, which is now also needed to give neutrinos a
mass. The 16 is also a fermionic representation. The beauty of the resulting scheme is that quarks,
antiquarks, leptons and antileptons all end up in one multiplet, which means that 3-quark systems can decay
into leptons, so the proton would become unstable. What I find particularly appealing about this construction
is that one ends up with a typically fermionic representation of SO(10), as if nature decided that if
something is a fermion in space-time, it better also be a fermion in internal space; thus, with 3
generations, we apparently have three fermions, each forming 32 dimensional (chiral) spinors in a 10+4
dimensional ``space-time".

Salam's picture of the natural world was that it should be described by physical laws that are aesthetically
pleasing. Beauty was a very important criterion for him. He was enchanted by supersymmetry, supergravity,
superspace and superstring theory. These theories had to be true just because they are beautiful. To my
taste, he got himself carried away sometimes a bit too easily. When a theory turns out to give unexpected new
insights, receiving support from different corners of experimental physics and/or theoretical arguments, it
usually also allows us a new view, showing us a new horizon that had been hidden up to that moment. Such new
horizons are always beautiful, so indeed, correct theories are beautiful theories, in general. But turning
this around very often does not work. It would have been perceived as ``beautiful" when matter was made of
just four elements, ``water", ``air", ``earth" and ``fire". This was the beauty that was searched for by the
primitive scientists of the Middle Ages; yet the real truth would turn out to be more beautiful in its own,
more complicated ways. This is a lesson that one should always keep in mind. There are beautiful worlds all
around us, but one might have to wade through a swamp to reach them.

\newsec{Superstrings and Quantum Gravity}

Injecting quantum mechanics into Einstein's theory of General Relativity turned out to be a vastly more
stubborn problem than most of us had anticipated. We thought that this was a problem very similar to the
riddles we have been facing in the past. Injecting quantum mechanics into Special Relativity had also been
hard, and it too at first looked like an impossible assignment. Various approaches using as much experimental
input as we could put our hands on, in combination with pure logical reasoning, were tried; and we
vindicated: the ``Standard Model" was the most precise and complete answer that was uncovered. So should we
not do the same thing again, sharpen our theoretical and experimental techniques, produce more precise
formalisms, and yes, quantum gravity will be there. Will it? History also shows us that procedures that have
proven very successful in the past, do not always guarantee success in the future. To crack this problem, we
might first have to step back. And if that does not help, perhaps step back further. By itself, this is a
risky thing to say. I receive letters every day from crackpots who give me the same advice: forget all that
mathematical gibberish, read the Bible, go meditate, or something of that sort. This is clearly not what I
have in mind.

What I do have in mind is that we have to improve our mathematics, but we might have to start at places where
we thought everything was secure. My own hobbyhorse is quantum mechanics. It is taken for granted by almost
all theoreticians that, in order to formulate some quantized version of gravity, one has to set up a
description of the basis elements of Hilbert Space. The first thing we teach to our students is that all
linear transformations in this Hilbert space are allowed, and all physical transformations of interest can be
reduced to such linear transformations in Hilbert space. There is no particular basis to be preferred in
favor of any other.

One might, however, suspect that quantum mechanics is not a completely immune corner stone of all fundamental
theories. It might be possible to \emph{explain} why we experience quantum mechanics in the world of the tiny
things. Such an explanation might reveal that there is something underneath quantum mechanics. The notion we
call Hilbert space is then just degraded into a powerful mathematical machine to handle the stochastic nature
of the solutions to some highly complex equations. Could we not try to identify candidates for such
underlying dynamical theories? I claim that we can indeed try to do this, and that this may well result is a
picture where one particular basis of Hilbert space actually describes ontological reality, whereas others do
not. This basis cannot be identified in terms of the particles and fields we know today, because probably the
complete set of all particle types, including the ones relevant at the Planck scale only, and the ones
describing black holes, have to be included. This will be a gargantuan task, but we can first try to find
simplified models where exactly this picture applies. One finds that this idea is not crazy, but we do have
to abandon some of our standard procedures. Since we cannot abandon all of our standard procedures at once,
this is a difficult path to follow, but it is worth thinking about it.

Compared to what I just said, superstring theory is just a more ``conventional" scheme. It is not my
intention to criticize superstring theory, but one cannot help noting that something essential appears to be
lacking: a precise description of a valid \emph{interpretation} of this theory. Looking at the face of
things, one is inclined towards the following interpretation.

What string theory adds to the construct of quantum field theories is not only a notion of one-dimensional -
linelike - structures, but also higher dimensional objects, (mem-)branes. There is a whole tower of
mathematical features that can be observed here, including general relativity in target space - i.e. gravity
in space-time. It is suggestive that this indeed is the structure that also plays a key role in General
Relativity. At the same time, however, there are numerous shortcomings if one wishes to elevate this
construct to the level of a fundamental ToE\fn{``Theory of Everything".}. The structure is fundamentally
quantum mechanical, which means that it will never make definite predictions, but only yield statistical
expectations for its fundamental variables. What is conspicuously missing is boundary conditions (in space as
well as in time), and variables that evolve deterministically. If one believes that Nature (or ``God") is
perfect, then this ``partial' theory is suspect.

The recent ``landscape" theories have had a major impact on our ``Grand View of Physics". The way the
situation is presently formulated appears to lead to a very disappointing state of affairs, since this scheme
is not much more than a gesture of surrender: we will never be able to derive the most conspicuous features
of the Standard Model. Yet the landscape idea is also difficult to refute. This really could be the
conclusion that our quest for understanding will be leading to. The power of superstring theory, together
with the impressive mathematical edifice that has been established by its investigators, should not be
underestimated. But as long as, in my view, this theory remains incomplete, requiring more solid foundations
than what we have at present, there may still be hope.

\end{document}